\documentclass[fleqn,usenatbib]{mnras}
\usepackage[T1]{fontenc}
\usepackage{lastpage}
\usepackage{ae,aecompl} 
\usepackage{graphicx}   
\usepackage[export]{adjustbox}
\usepackage{wrapfig}
\usepackage{amsmath}    
\usepackage{amssymb}  
\usepackage{comment}
\usepackage{cleveref}
\usepackage{natbib}
\defcitealias{Mineo_2012a}{M+12a}
\defcitealias{Fragos_2013a}{F+13}
\defcitealias{Madau_2017}{M+17}
\defcitealias{Brorby_2016}{B+16}
\defcitealias{Lehmer_19}{L+19}
\defcitealias{Fornasini_2019}{F+19}
\defcitealias{Fornasini_2020}{F+20}
\defcitealias{Yue_2015}{Y+15}
\defcitealias{Zahid_2014}{Z+14}
\defcitealias{Curti_2020}{C+20}
\defcitealias{Pallottini_2014}{P+14}
\defcitealias{Fontanot_21}{F+21}
\usepackage{hyperref}

\newcommand{\cmfast}{{\tt 21cmFAST}}
\newcommand{\cmmc}{{\tt 21cmMC}}
\newcommand{\LxSFR}{L_{\rm X}/{\rm SFR}}

\title[HMXBs metallicity and the cosmic 21-cm signal]{The 21-cm signal from the Cosmic Dawn: metallicity dependence of high mass X-ray binaries}
\author[Kaur et. al.]{Harman Deep Kaur$^{1}$\thanks{E-mail: harman.kaur@sns.it}, Yuxiang Qin$^{1, 2,3}$, Andrei Mesinger$^1$, Andrea Pallottini$^{1}$,\and Tassos Fragos$^4$, Antara Basu-Zych$^{5,6,7}$ \\
$^{1}$Scuola Normale Superiore, Piazza dei Cavalieri, 7 56126 Pisa, Italy\\
$^{2}$School of Physics, University of Melbourne, Parkville, VIC 3010, Australia\\
$^{3}$ARC Centre of Excellence for All Sky Astrophysics in 3 Dimensions (ASTRO 3D)\\
$^{4}$Département d’Astronomie, Université de Genève, Chemin Pegasi 51, CH-1290 Versoix, Switzerland\\
$^{5}$Department of Physics, University of Maryland Baltimore County, Baltimore, MD 21250, USA \\
$^{6}$NASA Goddard Space Flight Center, Laboratory for X-ray Astrophysics, Greenbelt, MD 20771, USA \\
$^{7}$Center for Research and Exploration in Space Science and Technology, NASA/GSFC, Greenbelt, MD 20771 }

\begin{document}

\maketitle

\begin{abstract}
X-rays from High-Mass X-ray Binaries (HMXBs) are likely the main source of heating of the intergalactic medium (IGM) during Cosmic Dawn (CD), before the completion of reionization.  This Epoch of Heating (EoH; $z\sim$10--15) should soon be detected via the redshifted 21-cm line from neutral hydrogen, allowing us to indirectly study the properties of HMXBs in the unseen, first galaxies.  Low-redshift observations, as well as theoretical models, imply that the integrated X-ray luminosity to star formation rate of HMXBs  ($\LxSFR$) should increase in metal-poor environments, typical of early galaxies.
Here we study the impact of the metallicity ($Z$) dependence of $\LxSFR$ during the EoH.   For our fiducial models, galaxies with star formation rates of order $10^{-3}$ -- $10^{-1}$ $M_\odot$ yr$^{-1}$ and metallicities of order $10^{-3}$ -- $10^{-2}$ $Z_\odot$ are the dominant contributors to the X-ray background (XRB) during this period. 
Different $\LxSFR$--$Z$ relations result in factors of $\sim$ 3 differences in these ranges, as well as in the mean IGM temperature and the large-scale 21-cm power, at a given redshift.
We compute mock 21-cm observations adopting as a baseline a 1000h integration with the upcoming Square Kilometer Array (SKA), for two different $\LxSFR$--$Z$ relations.  We perform inference on these mock observations using the common simplification of a constant $\LxSFR$, finding that constant $\LxSFR$ models can recover the IGM evolution of the more complicated $\LxSFR$--$Z$ simulations only during the EoH.
At $z<10$, where the typical galaxies are more polluted, constant $\LxSFR$ models over-predict the XRB and its relative contribution to the early stages of the reionization.
\end{abstract}


\section{Introduction}

High Mass X-ray Binaries (HMXBs) are expected to be the dominant source of heating in the early Universe, soon after the formation of the first galaxies 
(e.g. \citealt{Furlanetto_2006, Mcquinn_2012,Fragos_2013a, Paccuci_2014}).
Their X-rays have long mean free paths, and are thus able to penetrate deep into the intergalactic medium (IGM), at a time when reionization was still in its infancy ($z\sim$  10--20; \citealt{Mirabel_2011, Mcquinn_2012, Madau_2017, Eide2018}). As a result, X-rays drive large-scale IGM temperature fluctuations, during this so-called Epoch of Heating (EoH; e.g. \citealt{PF_2007, Santos2010, Mesinger_2011, Visbal2012, Paccuci_2014, Munoz2021}). The corresponding signal from the 21-cm line of neutral hydrogen is expected to be detectable by the ongoing Hydrogen Epoch of Reionization (HERA; e.g. \citealt{DeBoer17})\footnote{\url{https://reionization.org/}} and Square Kilometer Array (SKA)\footnote{\url{https://www.skatelescope.org/}} interferometers (for a recent review see, e.g., \citealt{Mesinger2019}).

Current theoretical models of the EoH are based on empirical scaling relations (e.g. \citealt{Lehmer_2010,Kaaret_2011,Mineo_2012a, Basu-Zych_2013,  Konstantinos_2020}) between the (population-averaged) X-ray luminosity ($L_{\rm X}$) and star formation rates (SFRs) of local galaxies: $L_{\rm X}/\rm SFR$. However, both theoretical models of HMXB evolution (e.g. \citealt{Linden_2010, Fragos_2013a}) and recent observations (e.g. \citealt{Basu-Zych_2013, Prestwich_2013,Douna_2015,Brorby_2016,Lehmer_2021}) suggest that this $L_{\rm X}/\rm SFR$ relation should have a strong dependence on metallicity. Theoretically we expect that weaker stellar winds resulting from lower metallicity environments would result in more X-ray luminous binaries, due to both a reduced expansion of the binary orbit and a reduced radial expansion of the companion which impacts when Roche-Lobe overflow occurs.   
This metallicity dependence can be important for the EoH, since the first galaxies are expected to form inside pristine gas with a rapidly-evolving metallicity (e.g. \citealt{Wise_2012,Xu_2013,Pallottini_2014,Jaacks_2019,Ucci2021}). The cosmic 21-cm signal at those redshifts can therefore be used to study the properties of the first generations of HMXBs (e.g. \citealt{Madau_2017, YQ_2020}). 

In this work we study the imprint of metallicity-dependent $L_{\rm x}/\rm SFR$ relations in the 21-cm signal from the Cosmic Dawn\footnote{The 21-cm signal depends on metallicity also through its impact on the stellar population (e.g. \citealt{Magg_2021,Munoz_2022}), as well as dust attenuation of UV photons (e.g. \citealt{Mirocha_2019}).  Here we focus on its role in setting the X-ray luminosity to SFR relation, deferring more focused studies to future work.}.
  Assuming a mass-metallicity relation, we compute the evolution of the 21-cm signal corresponding to two different scalings of $L_{\rm x}/\rm SFR$ with metallicity\footnote{Throughout we use ``metallicity'' to refer to the gas-phase metallicity, $12+ \rm log_{10}(O/H)$. Solar gas-phase metallicity ($12+\log_{10}(\mathrm{O/H})_{\odot}$) is taken to be 8.69 (\citealt{Asplund_2004}). Absolute metallicity, $Z_{\odot}$ is taken to be 0.02.}.  We quantify how these relations impact the X-ray emissivity and IGM temperature evolution, and also make forecasts for the corresponding 21-cm signal.
 
This paper is organized as follows: In \S \ref{sec:xray}, we present the steps for computing the X-ray emissivity during the cosmic dawn. In \S \ref{sec:Thermal_evol}, we compute the associated thermal evolution of the IGM, followed by the  21-cm signal in \S \ref{sec:21cm_signal}. In \S \ref{sec:mcmc}, we quantify if simpler, constant $\LxSFR$ models can recover the same IGM evolution as predicted from the fiducial $L_{\rm X}/\rm SFR$--$Z$ relations. Finally we conclude in \S \ref{sec:Conclusions}. 
We assume a standard $\Lambda$CDM cosmology with the following cosmological parameters:
$
h =0.678 , \Omega_{\rm m} =0.308 , \Omega_{\rm b} =0.0484 , \Omega_{\Lambda} =0.692 ,\sigma_{8} = 0.815, n_{\rm s} = 0.968$ based on result from \citealt{2016A&A...594A..13P}.

\section{X-ray emission from early galaxies}
\label{sec:xray}

In this section we present our model for computing the inhomogeneous X-ray background during the Cosmic Dawn.  We begin by discussing the empirical and theoretical $L_{\rm X}/\rm SFR-Z$ relations we use in \S \ref{ssec:Lxsfr}.  We then show the adopted mass-metallicity relation for assigning metallicities to dark matter halos at high redshifts in \S \ref{ssec:MZR}. Finally in \S \ref{ssec:XRB} we compute the corresponding X-ray backgrounds by integrating over the halo mass functions and  star formation rates of galaxies.

To compute the metallicity-dependent X-ray backgrounds and their impact on the IGM, we modify the public code, \cmfast\ \footnote{\url{https://github.com/21cmfast/21cmFAST}}. \cmfast\  uses a combination of perturbation theory, excursion-set formalism and light-cone integration to compute 3D realizations of the IGM density, velocity, ionization, temperature and 21-cm intensity fields.  A detailed description is available in  \citealt{Mesinger_2007,Mesinger_2011, Murray_2020}.
Our simulations are 500 comoving Mpc per side, using a 256$^3$ grid smoothed down from $1024^3$ initial conditions.  These volumes are large enough to accurately sample cosmic variance during the EoH (\citealt{Kaur_2020}; Balu et al. in prep).

\subsection{Metallicity dependent X-ray luminosity to SFR relations}
\label{ssec:Lxsfr}
We assume that early galaxy populations can be characterized by an average $L_{\rm X}/\rm SFR $ relation.
This relation is often taken to be a constant when computing the EoH and associated 21-cm signal (though see, e.g. \citealt{Madau_2017, Eide2018}).  Here we explore how the 21-cm signal is impacted by the metallicity dependence of this relation, assuming two very different $L_{\rm X}/\rm SFR$ -- $Z$ scalings.

The first is taken from \citet{Brorby_2016} (hereafter,  \citetalias{Brorby_2016}), who studied local, metal-poor, star-forming galaxies obtaining:
\begin{equation}
\label{eq:BrorbyLx}
\begin{split}
        \log_{10}
        \left(\frac{L_{\rm X,[0.5-8\ keV]}}{\rm erg\ s^{-1}}\right) & = a\rm \log_{10} \left(\frac{\rm SFR}{\rm{M_{\odot}}\ \rm{yr^{-1}}}\right) \\
        &  + b\rm \ \log_{10}\left(\frac{\rm (O/H)}{(\mathrm{O/H})_{\odot}}\right) + \it c,
\end{split}  \end{equation}
where $L_{\rm X,[0.5-8\ keV]}$ is the X-ray luminosity in $0.5-8\ \rm keV$ range, $\rm a = 1.03 \pm 0.06, \rm b = - 0.64 \pm 0.17, \rm  c = 39.46 \pm 0.11$.

\begin{figure}
     \includegraphics[width=0.5\textwidth]{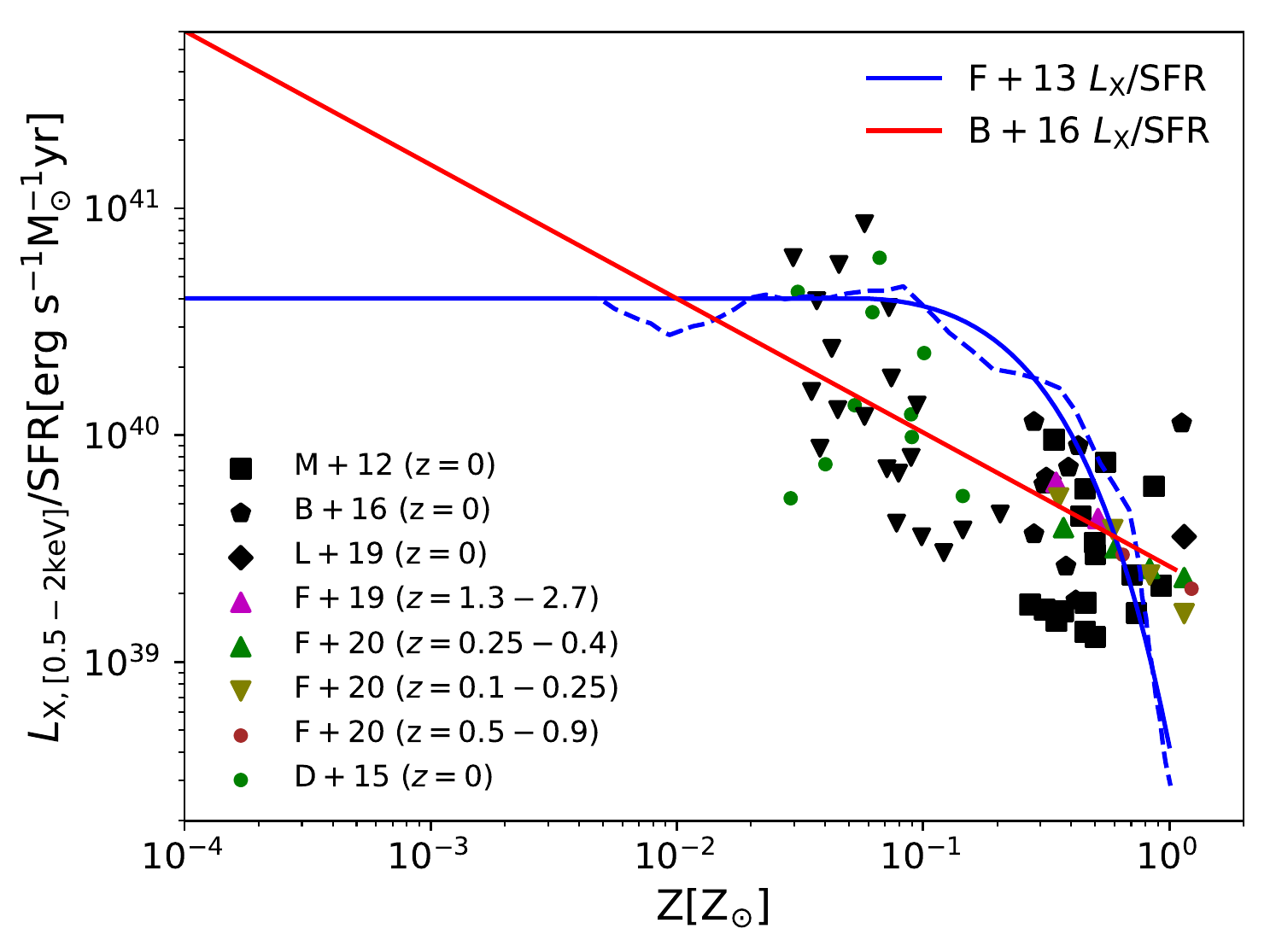}
    \caption{Soft-band $L_{\rm X}/\rm SFR$ vs $Z$. The \citetalias{Fragos_2013a} curve (solid blue) is constructed by fitting the XRB synthesis model 265 (dashed blue) of \citealt{Fragos_2013a}.  The red line is the empirical fit of observation data of local Lyman break analogs (black pentagons) from \citealt{Brorby_2016}. To compare these models with observations, we also plot some observational estimates from the literature.  Black squares are $L_{\rm X}/\rm SFR$ estimates of $z=0$ HMXB dominant star-forming galaxies from \citet{Mineo_2012a}, constructed using X-ray, infrared and UV data from Chandra, SPITZER, GALEX, and 2MASS archives. Their corresponding metallicities were taken from \citet{Douna_2015}. Black diamonds are resolved samples of galaxies observed in both infrared (SINGS survey) and Chandra X-ray from \citet{Lehmer_19}. Green circles correspond are from \citet{Douna_2015}, estimated primarily from blue compact dwarf galaxies, with upper limits denoted by upside-down triangles in black. Stacked Chandra data from the COSMOS Legacy survey (\citealt{Fornasini_2019,Fornasini_2020}) in the redshift range $z=  0.1 - 2.7$ is shown with corresponding triangles. All data points are converted into the soft-band.} 
    \label{fig:Lxsfr}
\end{figure}

The second $L_{\rm X}/\rm SFR$ -- $Z$ relation is taken from stellar evolution models of HMXBs \citep{Fragos_2013a}.   
Specifically, we use model 265 (hereafter, \citetalias{Fragos_2013a}) which is also the maximum likelihood model in \citet{Madau_2017}. 
We find that the following Schechter function provides a good fit for its $L_{\rm X}/\rm SFR$ -- $Z$ relation:
\begin{equation}
\log_{10}\left(\frac{\it L_{\rm bol}/\rm SFR}{\rm erg\ s^{-1}\ M_{\odot}^{-1}yr}\right) = A + \alpha \log_{10}\left(Z/Z_{\rm turn}\right) - Z/Z_{\rm turn}
\label{eq:FragosLx}
\end{equation}
Here, $\it L_{\rm bol}/\rm SFR$ is the bolometric X-ray luminosity per star formation rate in $\rm erg\  s^{-1} M_{\odot}^{-1}\rm yr$, $A= 41.3, \alpha = 0.3,\ \text{and} \ Z_{\rm turn} = 8\times 10^{-3}$. We also set the X-ray luminosity to constant below the peak value at $Z/Z_{\rm turn}\le \alpha / \ln 10$, to prevent the empirical fit from decreasing towards the lowest metallicities, in agreement with the constant $\LxSFR$ predicted by the \citetalias{Fragos_2013a} model at these metallicities.

We show the \citetalias{Brorby_2016} and \citetalias{Fragos_2013a} relations in Figure \ref{fig:Lxsfr}, together with some observational data (see caption for details).  Since only soft X-rays are efficient in heating the IGM, we convert all luminosities into the soft-band (0.5 -- 2 keV), using the two component plus a power-law model from \citet{Mineo_2012a}.  We note that observations based on individual galaxies show significant scatter, especially at low SFRs (roughly corresponding to low metallicities).  
This is due to sparse sampling of the relatively broad luminosity functions (LFs) of X-ray binaries, which require $\sim$ tens of galaxies to get unbiased estimates of the population-average values of $\LxSFR$ (e.g. \citealt{Lehmer_2021}).

From the figure, we see that both \citetalias{Brorby_2016} and \citetalias{Fragos_2013a} $L_{\rm X}/{\rm SFR}$ -- $Z$ relations pass through the observational data; however, they show dramatically different trends towards very low metallicities.  As we shall see further below, galaxies with $Z \lesssim 10^{-2} Z_\odot$ likely dominate the X-ray background during the EoH.  In this regime, the model from \citetalias{Fragos_2013a} asymptotes to a much lower value of $L_{\rm X}/{\rm SFR}$ than the empirical relation from \citetalias{Brorby_2016}. The theoretical explanation for this is that once metallicity goes below a certain value (roughly 0.05 $Z_\odot$), stellar winds become inefficient and only have a marginal impact on the stellar evolution (e.g. \citealt{Fragos_2013a}).  However, these models do not vary the initial mass function (IMF), which could have an additional impact on the HMXB luminosity at very low metallicities (c.f. Fig. 7 in \citetalias{Fragos_2013a}).
Moreover, we note that the \citetalias{Brorby_2016} $\LxSFR$--$Z$ relation is derived empirically, and was not intended to be extrapolated to the lowest metallicities we show in the figure.

Therefore, the $\LxSFR$--$Z$ relation is highly uncertain at the lowest metallicities.  It is important to note however that here we are not arguing for the validity of any specific $\LxSFR$ -- $Z$ relation.  Instead, we are exploring the observational impact of $\LxSFR$ -- $Z$ relations during the Cosmic Dawn. Therefore the \citetalias{Brorby_2016} and \citetalias{Fragos_2013a} scalings are useful in that they assign very different properties to the first generations of HMXBs.  Indeed, we would want to eventually {\it infer these relations directly from future 21-cm data.}

\subsection{Mass-metallicity relation}
\label{ssec:MZR}

In the previous section we presented the two different $\LxSFR$--$Z$ relations we  use in this study.  We now discuss our adopted prescription for assigning $Z$ to high-redshift galaxies.

The galaxy-averaged gas-phase metallicity shows a strong correlation with the galaxy's stellar mass ($M_{*}$), commonly referred to as the mass-metallicity relation (MZR; \citealt{Mannucci_2010,Lara_Lopez_2010,Yates_2012,Zahid_2014,Cresci_2019,Curti_2020})\footnote{The mass-metallicity relation can also be understood as the 2D projection of the more general 3D fundamental metallicity relation (FMR), relating metallicity to the stellar mass and SFR (e.g. \citealt{Mannucci_2010, Hunt_2012}).  In our analytic model, the stellar mass and SFR are themselves related deterministically (i.e. through the main sequence of star forming galaxies), thus it suffices to only specify one or the other.}. It has been observed in the local Universe and even at redshifts as high as $z =3.5$ (e.g., \citealt{Tremonti_2004,Erb_2006,2008A&A...488..463M,Ellison_2008,Mannucci_2010,Zahid_2014,2018ApJ...858...99S,Curti_2020,Sanders_2021}), and has also been found in theoretical works (e.g., \citealt{Yates_2012, Ucci2021}). 

Here we adopt the empirical MZR from \citet{Zahid_2014}:
\begin{equation}
Z = Z_{0} + \rm \log_{10}(1 - e^{-(\it M_{*}/M_ {\rm 0})^{\gamma}}) ,
\label{eq:Zahid}
\end{equation}
where $Z_{0}$ is the saturation metallicity, $M_{0}$ is a characteristic stellar mass scale above which the metallicity asymptotically approaches $Z_{0}$. The relation reduces to a power-law with index $\gamma$ at $M_{*}$ < $M_{0}$, and flattens at higher masses. The redshift dependence is incorporated in $M_{0} \equiv 10^b$.
By fitting to data from the Sloan Digital Sky Survey (SDSS), the Deep  Extragalactic Evolutionary Probe 2 (DEEP2), FMOS-COSMOS, and Smithsonian Hectospec Lensing Survey (SHELS), that span a redshift range up to 1.6, \citet{Zahid_2014} find a set of best-fit parameters as follows:
$\rm Z_0 = 9.100$, $b = 9.135 + 2.64\log_{10}(1 + z)$ and $\gamma = 0.522$.

\begin{figure}
\centering
\includegraphics[width=0.5\textwidth]{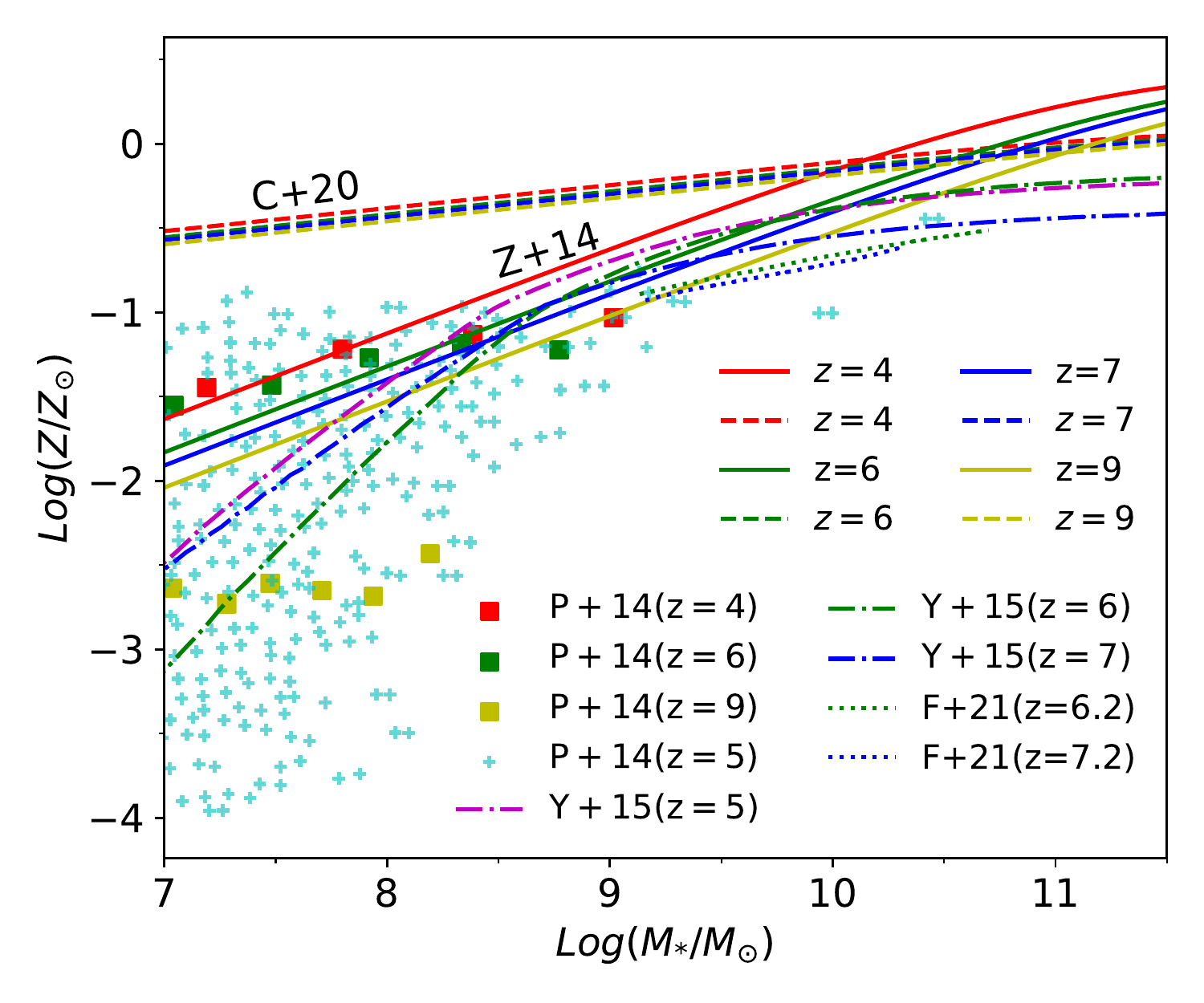}
\caption{Gas-phase metallicity vs stellar mass relation (MZR) from \citet{Zahid_2014} (eq. \ref{eq:Zahid}) is shown with solid curves (Z+14) and from \citet{Curti_2020} with dashed curves (C+14). The curves are color-coded according to redshift. We also plot mean relations from the hydrodynamic simulations of \citet{Pallottini_2014} (P+14, squares), \citet{Yue_2015} (Y+15, dot-dashed curves), and semi-analytic models of \citet{Fontanot_21} (F+21, dotted curves). Cyan crosses denote values from individual galaxies in \citet{Pallottini_2014} at $z=5$, illustrating the galaxy-to-galaxy scatter. }
\label{fig:Mass_metallicity}
\end{figure}

We illustrate this relation in Figure \ref{fig:Mass_metallicity} at $z=4$--9 using solid curves. For reference, we also show data from the cosmological, hydrodynamic simulations of \citet{Pallottini_2014}, in which they follow interstellar medium (ISM) and IGM metal enrichment. \citetalias{Yue_2015} curves are from \citet{Yue_2015} where they use the stellar mass to UV magnitude relation from \citet{Duncan_2014}, combined with the FMR from \citet{Mannucci_2010}. We also show $Z$--$M_{*}$ predictions from semi-analytical GAEA models of \citet{Fontanot_21}, tested using the VANDELS survey (\citealt{McLure_2018,Pentericci_2018}).  
Additionally, we show the empirical MZR from \citet{Curti_2020}, which depends on both stellar mass and SFR. This relation is more similar to the FMR (\citealt{Mannucci_2010}), which is usually regarded as redshift-independent.  In making this comparison figure, we used the stellar -- halo mass and the SFR -- stellar mass relations from \citet{Park_2019} when required; these were calibrated to reproduce high-redshift UV LFs and other EoR data (see \S \ref{ssec:XRB} for more details).

From Figure \ref{fig:Mass_metallicity} we see that the high-redshift simulations and semi-analytic models are in reasonable agreement with the MZR from \citet{Zahid_2014}, though the evolution towards the highest redshifts and smallest masses is highly uncertain.  Therefore the 21-cm forecasts we present in \S \ref{sec:mcmc} should be taken with caution.  

\subsection{X-ray emissivity during the Cosmic Dawn}
\label{ssec:XRB}

The $\LxSFR$--$Z$ relations and MZR discussed in the previous sections allow us to compute an average X-ray luminosity for a galaxy, given its SFR and stellar mass.  To compute the X-ray background we then need to relate the SFR and stellar mass to the typical mass of the host halo, since then we can integrate over well-established halo mass functions (HMFs).
Here we use the simple and flexible power law scaling relations from \citet{Park_2019}, shown to recover various reionization era observables.
  Specifically, we assume a power law for the stellar to halo mass relation:
\begin{equation}\label{eq:F_STAR}
\frac{M_{\ast}}{M_{\rm h}} = f_{\ast,10} \left( \frac{M_{\rm h}} {10^{10}{\rm M}_{\sun}} \right)^{\alpha_{\ast}} \left(\frac{\Omega_{\rm b}}{\Omega_{\rm m}}\right),
\end{equation}
where  $f_{\rm \ast}(M_{\rm h})=f_{\rm \ast,10}(M_{\rm h}/10^{10} {\rm M}_\odot)^{\alpha_{\ast}}$ is the fraction of the galactic gas in stars and $\alpha_{\ast}$ is the power-law index.  Similarly, we relate the fraction of UV ionizing photons which escape into the IGM, $f_{\rm esc}$ to halo mass with: $f_{\rm esc}(M_{\rm h})=f_{\rm esc,10}(M_{\rm h}/10^{10} { M}_\odot)^{\alpha_{\rm esc}}$.  The SFR can be expressed as:
\begin{equation}\label{eq:SFR}
SFR = \dot{M}_\ast =  \frac{M_{\ast}}{t_\ast H(z)^{-1}}, 
\end{equation}
where $t_{\ast}$ is a typical star formation timescale as a fraction of the Hubble time, $H(z)^{-1}$, with $H(z)$ being the Hubble parameter at $z$.  
These simple scaling relations seem sufficient to characterize the population-averaged properties of high-$z$ galaxies.  They are consistent with UV LFs at $z=6$--10 and $M_{\rm uv} >= -20$ (e.g. \citealt{Bouwens_2014,Bouwens_2016, McLeod_2016, Finkelstein_2016,Livermore_2017,Oesch_2018, Gillet_2020,Atek_2018,Ishigaki_2018,Bouwens_2021})\footnote{Here we are only interested in the faint-end galaxies that dominate the photon budget (e.g \citealt{YQ_2021}).  We therefore do not include a separate power law at the bright end, typically associated with AGN feedback (e.g. \citealt{Furlanetto_2006,Behroozi_2015,Mirocha_2017, Sabti_2021,Rudakovskyi_2021}).  Such bright galaxies sit on the exponential tail of the HMF, and are thus too rare to be important for determining cosmic radiation fields.}.  Moreover, both hydrodynamical simulations (e.g. \citealt{Xu_2016,Ma_2020}) and semi-analytic models (e.g. \citealt{Sun_2016,Mutch_2016,Behroozi_2019}) also imply power laws for the stellar-to-halo mass relation.

With the above, we can express the comoving, soft-band X-ray emissivity (in $\rm erg\ s^{-1} \ Mpc^{-3}$) at position ${\bf x}$ and redshift $z$ as:
\begin{equation}
\epsilon_{\rm 0.5-2 keV}({\bf x}, z) = 
\int_{0}^{\infty} \rm d \it M_{\rm h} \frac{\rm d\it n}{\rm d\it M_{\rm h}} e^{-\left(\frac{M_{\rm turn}}{M_{\rm h}}\right)} \dot{M}_\ast \frac{L_{\rm X}}{\rm SFR} ~.
\label{eq:equation_emissivity}
\end{equation}
Here ${\rm d\it n}/{\rm d\it M_{\rm h}}$ is the local (conditional) halo mass function, $e^{-(M_{\rm turn}/M_{\rm h})}$ accounts for the suppression in star formation in halos below a characteristic mass scale (i.e. $M_{\rm turn}$) due to inefficient cooling and/or feedback (e.g. \citealt{Hui_1997,Barkana_2001,Springel_2003,Mesinger_2008}), and $\LxSFR (Z, M_\ast, z)$ is related to the halo mass through equations (1) -- (5). 
 When computing mock observations below, we use these fiducial values:
$f_{*,10}$ = 0.05, \
$\alpha_*$ = 0.5,  \
$f_{\rm esc,10}$ = 0.1, \
$\alpha_{\rm esc}$ = -0.5, \
$M_{\rm turn}$ = $5\times 10^{8} \rm M_{\odot} $ and
$t_*$ = 0.5. These values correspond to the maximum a posteriori model in \citet{Park_2019}, which used UV LFs, the electron scattering optical depth to CMB, and the QSO forest dark fraction in the likelihood.

\begin{figure}
   \centering
   \includegraphics[width=0.46\textwidth]{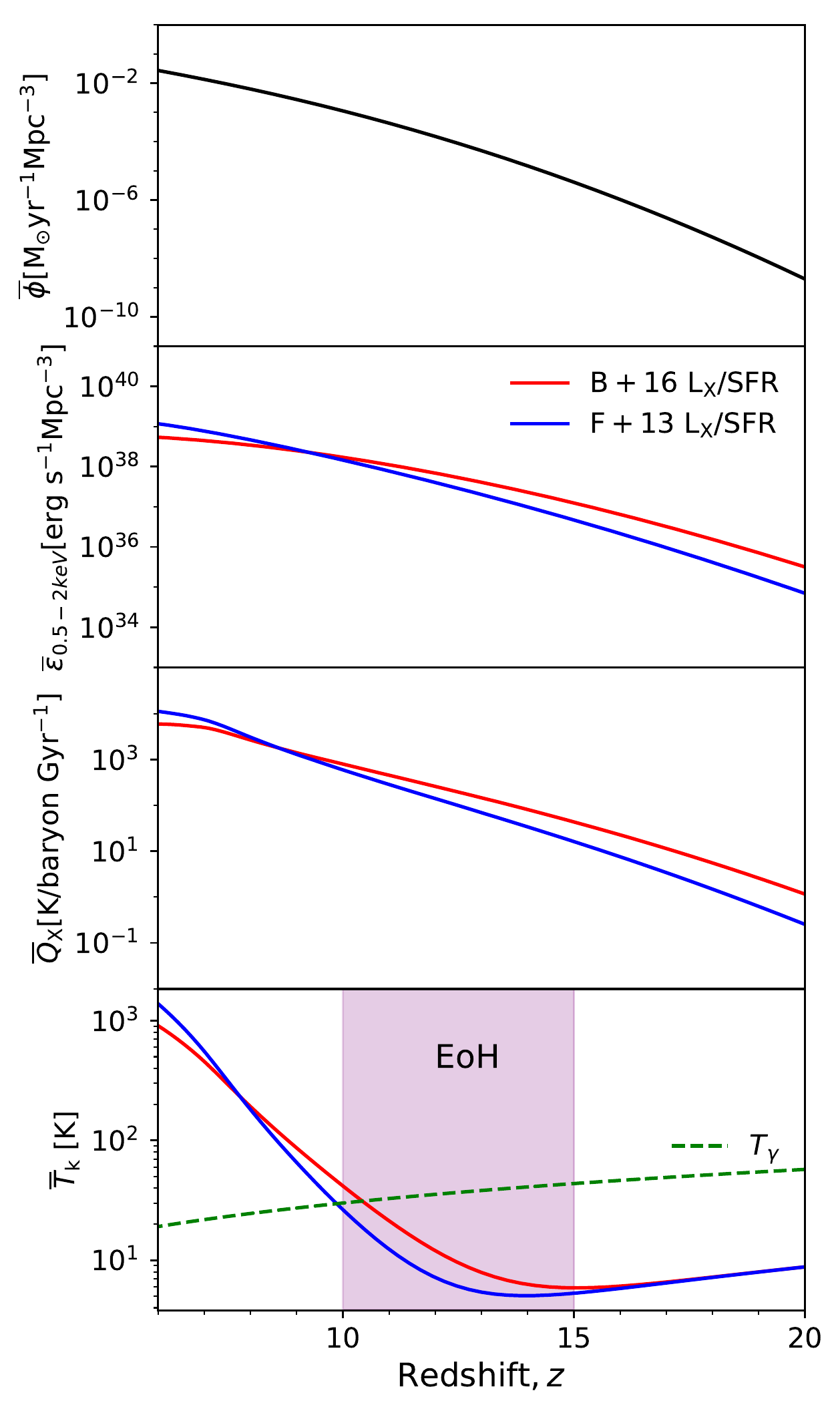}
 \caption{{\it From top to bottom}: redshift evolution of the (volume-averaged) star formation rate density (SFRD), soft-band X-ray emissivity (eq. \ref{eq:equation_emissivity}),  X-ray heating rate per baryon, $\overline{Q}_{\rm X}$ (eq. \ref{eq:heating_rate}), and kinetic temperature of the neutral IGM, $\overline{T}_{\rm k}$ (eq. \ref{eq:Tk}) computed using \cmfast. Red and blue curves correspond to the $L_{\rm X}/\rm SFR$--$Z$ relations from \citetalias{Brorby_2016} and \citetalias{Fragos_2013a}, respectively. In the bottom panel we also show the cosmic microwave background (CMB) temperature ($T_{\gamma}$) evolution with the dashed line, and approximately demarcate the Epoch of Heating (EoH).}
 \label{fig:Emissivities}
\end{figure}

In the top panel of Figure \ref{fig:Emissivities} we show our fiducial star formation rate density (SFRD) evolution, with the corresponding X-ray emissivity shown in the second panel (for both \citetalias{Brorby_2016} and \citetalias{Fragos_2013a} $\LxSFR$--$Z$ relations).  Because structure formation is hierarchical, the mass and metallicity of the typical galaxy population increase with time.  At early times, most galaxies were extremely metal-poor and so the \citetalias{Brorby_2016} $L_{\rm X}/\rm SFR$--$Z$ relation implies an X-ray emissivity that is a factor of $\sim$5 larger than the one from \citetalias{Fragos_2013a} at $z\sim20$.  When the characteristic metallicity of star-forming galaxies surpasses $Z\gtrsim0.01 Z_\odot$, the \citetalias{Fragos_2013a} relation implies higher X-ray luminosities than \citetalias{Brorby_2016} (see Figure \ref{fig:Lxsfr}).  From the second panel of Figure \ref{fig:Emissivities}, we see this transition happening at $z\sim10$, with \citetalias{Fragos_2013a} implying a higher emissivity over the range $6 < z < 10$.  As can be seen from Figure 1, we expect this trend to reverse again, with \citetalias{Brorby_2016} having higher luminosities at $Z \gtrsim Z_\odot$.  However, such highly metal-enriched galaxies are too massive and rare to be relevant during the Cosmic Dawn.

\begin{figure}
    \centering
    \includegraphics[width=0.5\textwidth]{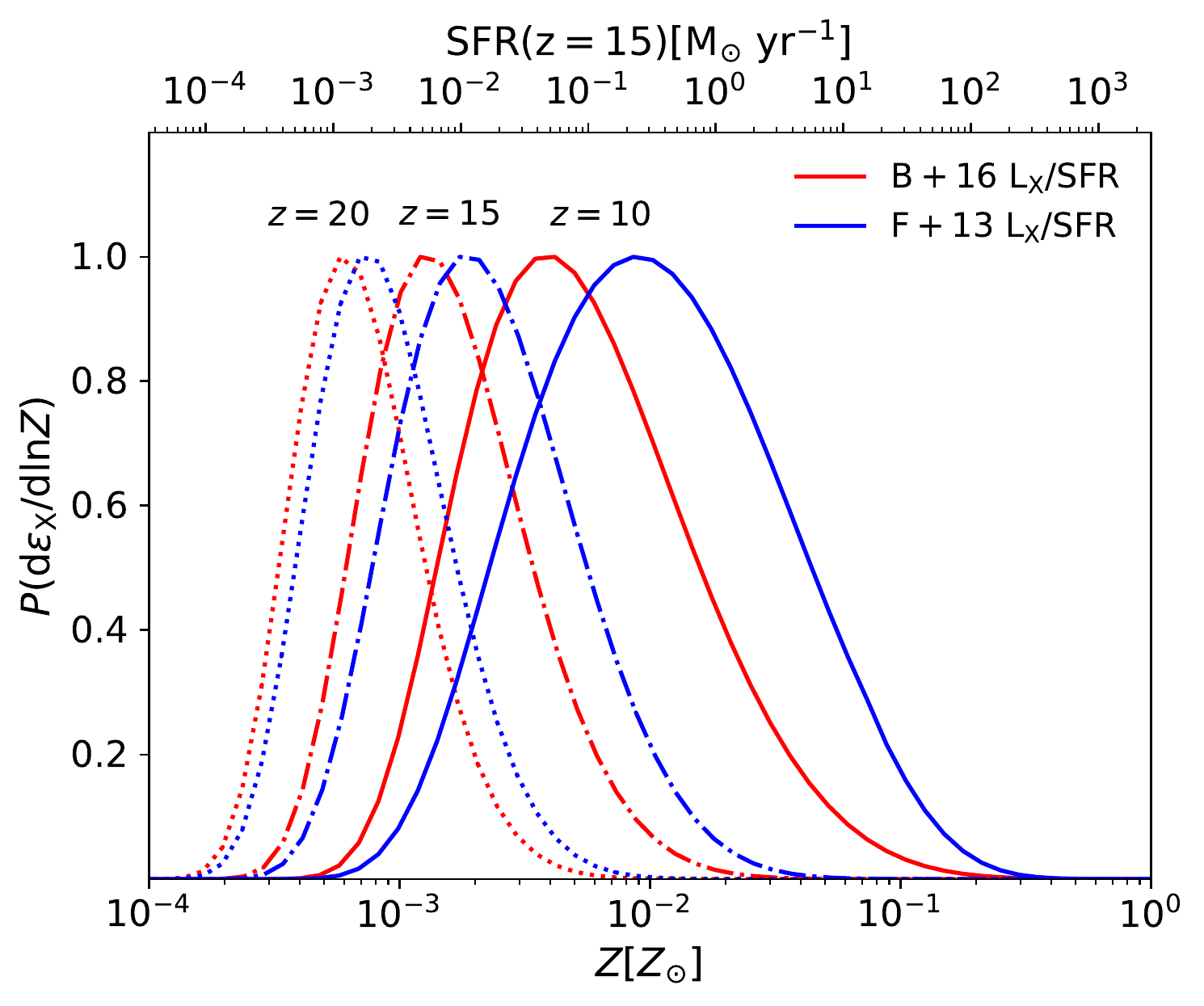}
    \caption{
Fractional contribution to the total X-ray emissivity sourced by galaxies per logarithmic metallicity bin. \citetalias{Brorby_2016} and \citetalias{Fragos_2013a} $L_{\rm X}/\rm SFR$--$Z$ relations are shown with red and blue curves respectively. Here different line-styles represent different redshifts. On the top x-axis, we denote the corresponding SFR at $z=15$.  For viewing purposes, we normalize all PDFs to peak at unity.}
\label{fig:Emissivity_function}
\end{figure}

We further quantify properties of the relevant galaxies in Figure \ref{fig:Emissivity_function}, where we plot the fractional contribution to the X-ray emissivity of galaxies within a given logarithmic metallicity bin.  The probability distribution functions (PDFs) shift to higher metallicities  and become wider with decreasing redshift.  This is driven by the evolution of the HMF which shifts to larger masses and flattens with time.
We see explicitly that the transition when the \citetalias{Fragos_2013a} emissivity surpasses that of \citetalias{Brorby_2016} at $z \lesssim 10$ corresponds to when the mean of the PDFs goes above $Z\sim 0.01 Z_\odot$.  As we discuss further below, the EoH for these models corresponds to $10 \lesssim z \lesssim 15$.  As a result, the EoH is driven by galaxies with metallicities $10^{-3} \lesssim Z/Z_\odot \lesssim 10^{-2}$ and SFRs roughly in the range $10^{-3} \lesssim \dot{M_\ast}/M_\odot \rm yr^{-1} \lesssim 10^{-1}$ (c.f. the top axis shows the corresponding SFRs implied by our model at $z=15$).

\section{Evolution of IGM properties}

\subsection{Temperature}
\label{sec:Thermal_evol}

We now compute the global evolution of the thermal and ionization state of the IGM, for the two X-ray emissivities shown in the previous section.
We can write the X-ray specific intensity, $J$ (in units of $\rm erg\ s^{-1}\ keV^{-1}\ cm^{-2} \ sr^{-1}$) for photons with energy $E_{\rm X}$, as seen by a gas element at position $\mathbf{x}$ and redshift $z$, as :
\begin{equation}
    J(\mathbf{x}, z, E_{\rm X}) = \frac{(1+z)^3}{4\pi}\int_{z}^{\infty}\rm d\it z' \frac{c\rm d\it t}{\rm d\it z'}\epsilon_{\nu} e^{-\tau} ~.
\label{eq:Jx}  \end{equation}
Here $\epsilon_{\nu}({\bf x}, z')$ is the {\it specific} X-ray emissivity, evaluated in the rest frame $E_0=E_{\rm X}(1+z')/(1+z)$, assuming a power-law spectral energy distribution (SED) with energy index $\alpha_{\rm X}=1$ and integrated soft-band value according to eq. \ref{eq:equation_emissivity}. The term $\exp[-\tau (z, z', E_{\rm X})]$  accounts for attenuation by hydrogen and helium through a two-phased IGM between $z$ and $z'$ (see \citet{Mesinger_2011} for more details).

This X-ray intensity drives the evolution of the ionized fraction ($x_{\rm e}$) and kinetic temperature ($T_{\rm k}$) of the (mostly neutral) IGM gas outside the HII regions proximate to galaxies according to:
\begin{equation}
\frac{\textup{d} x_{\rm e}(\mathbf{x},z)}{\textup{d} z} = \frac{\textup{d} t}{\textup{d} z} \left(\Gamma_{\rm X} - \alpha_{\rm A}C x_{\rm e}^{2}n_{\rm b}f_{\rm H}  \right)
\end{equation}
and
\begin{equation}
\begin{split}
    \frac{\textup{d} T_{\rm k}(\mathbf{x},z)}{\textup{d} z} &= \frac{2}{3k_{\rm b}(1+x_{\rm e})}\frac{\textup{d} t}{\textup{d} z}\sum Q_{\rm p} 
   +
   \frac{2T_{\rm k}}{3n_{\rm b}} \frac{ \textup{d} n_{\rm b}}{\textup{d} z} \\
   & - \frac{T_{\rm k}}{1+x_{\rm e}} \frac{ \textup{d} x_{\rm e}}{\textup{d} z} ~,
    \label{eq:Tk}
\end{split}    
\end{equation}
where $n_{\rm b}$ is the baryon number density, $ \Gamma_{\rm X}$ is the ionization rate per baryon from X-rays, $\alpha_{\rm A}$ is the case-A recombination coefficient, $C=\langle n_{\rm H}^2\rangle/\langle n_{\rm H}\rangle^2$ is the clumping factor on the scale of the simulation cell, $n_{\rm H}$ is the hydrogen number density, $f_{\rm H} $ is the hydrogen number fraction, $k_{\rm b}$ is the Boltzmann constant, and the radiative heating rate per baryon, $Q_{\rm p}$, includes both Compton heating and X-ray heating.
The X-ray heating and ionization rates can be expressed as:
\begin{equation}
    Q_{\rm X}(\mathbf{x},z) = \int
    d\nu \dfrac{4\pi J}{h\nu}\sum_{i}(h\nu - E_{i}^{\rm th})f_{\rm heat}f_{i}x_{i}\sigma_{i}
\label{eq:heating_rate}  \end{equation}
\begin{equation}
    \Gamma_{\rm X}(\mathbf{x},z) = \int
    d\nu \dfrac{4\pi J}{h\nu}\sum_{i} f_{i} x_{i} \sigma_{i} F_{i} 
\end{equation}
 where
\begin{align}
      F_{i} = (h\nu - E_{i}^{th})\left(\frac{f_{\rm ion,HI}}{E_{\rm HI}^{\rm th}} + \frac{f_{\rm ion,HeI}}{E_{\rm HeI}^{\rm th}} + \frac{f_{\rm ion,HeII}}{E_{\rm HeII}^{\rm th}}  \right) + 1
\end{align}
Here i stands for the atomic species: H, HeI and HeII, $E_i^{\rm th}$ is their corresponding ionization threshold, $f_{i}$ their number fraction, $x_{i}$ the ionization fraction (which for HI and HeI we take to be $x_{i} = (1-x_e)$ and for HeII $x_i = x_e$), $\sigma_i$ is the photo-ionization cross-section, $f_{\rm heat}$ is the fraction of the primary ionized electron's energy dissipating as heat and $f_{\rm ion,\it j}$ is its energy contributing to secondary ionization of the species \it j, \rm taken from \citet{FS10}.
Following the ISM simulations discussed in \citet{Das_2017}, we assume photons with energies below $E_0=0.5$ keV are absorbed by the host galaxies and unable to escape into the IGM.

In the bottom two panels of Figure  \ref{fig:Emissivities}, we show the volume-averaged $Q_{\rm X}$ and  $ T_{\rm k}$ corresponding to our two fiducial $L_{\rm X}/\rm SFR$--$Z$ relations. In the temperature panel, we also show the evolution of the CMB temperature, that provides the radio background in standard 21-cm models.  We roughly demarcate the EoH as shown in the figure.

The heating rate and temperature follow the same qualitative trend seen in the emissivity panel, with \citetalias{Brorby_2016} resulting in a higher temperature at redshifts $z\gtrsim8$.  The transition redshift at which \citetalias{Fragos_2013a} surpasses \citetalias{Brorby_2016} is somewhat lower for temperature (bottom panel) compared to the emissivity (second panel).  This is because the temperature depends on the light-cone integral over the emissivity, and not just its instantaneous value.  The flattening of the heating rate seen at $z \lesssim8$ is due to reionization by UV photons, which decreases the neutral fractions of hydrogen and helium.

\subsection{21-cm signal}
\label{sec:21cm_signal}

 As mentioned in the introduction, we use the cosmic 21-cm signal as our fiducial observational data-set, as it has the largest potential of constraining the thermal state of the gas in those early epochs.  Indeed, the recent HERA season 1 observations \citep{HERA_2021a} already constrain the $\LxSFR$ of high-$z$ galaxies, disfavoring the local $\LxSFR$ relation by > $1 \sigma$ \citep{HERA_2021b}. As discussed above, this is consistent with expectations that the first sources were more X-ray efficient due to their lower metallicities.

 The 21-cm signal can be expressed as the offset of the 21-cm brightness temperature ($T_{\rm b}$) from the CMB temperature ($T_{\gamma}$) at the observed frequency $\nu$ (\citealt{Furlanetto_2006}):
 \begin{align}
      \nonumber \delta T_{\rm b}(\nu) &= \frac{T_{\rm S}- T_{\gamma}}{1+z}(1-e^{-\tau_{21}}) \\
      &\approx 27x_{\rm HI}(1+\delta)\left(\frac{H}{\textup{d}v_{r}/\textup{d}r+H}   \right)\left(1- \frac{T_{\gamma}}{T_{\rm S}}\right) \nonumber \\
     \times & \left(\frac{1+z}{10} \frac{0.15}{\Omega_{\rm m} h^2}\right)^{1/2}\left(\frac{\Omega_{\rm b} h^2}{0.023}\right) \ \rm mK .
    \label{eq:delT}
\end{align}
\noindent Here, $\tau_{21}$ is the optical depth at the 21-cm frequency, $\delta$ is the gas over-density ($\delta = \rho/\overline{\rho}-1$), $\textup{d}v_{r}/\textup{d}r$ is the peculiar velocity gradient along the line-of-sight, and $T_{\rm S}$ is the spin temperature defined by the relative abundances of the excited and ground states of the hyperfine level of neutral hydrogen atom. Although \cmfast\ solves for the exact expression in the first row of eq. \ref{eq:delT}, we show the $\tau_{21} \ll 1$ approximation in the second row for physical intuition.

\begin{figure}
     \centering
    \includegraphics[width=0.5\textwidth]{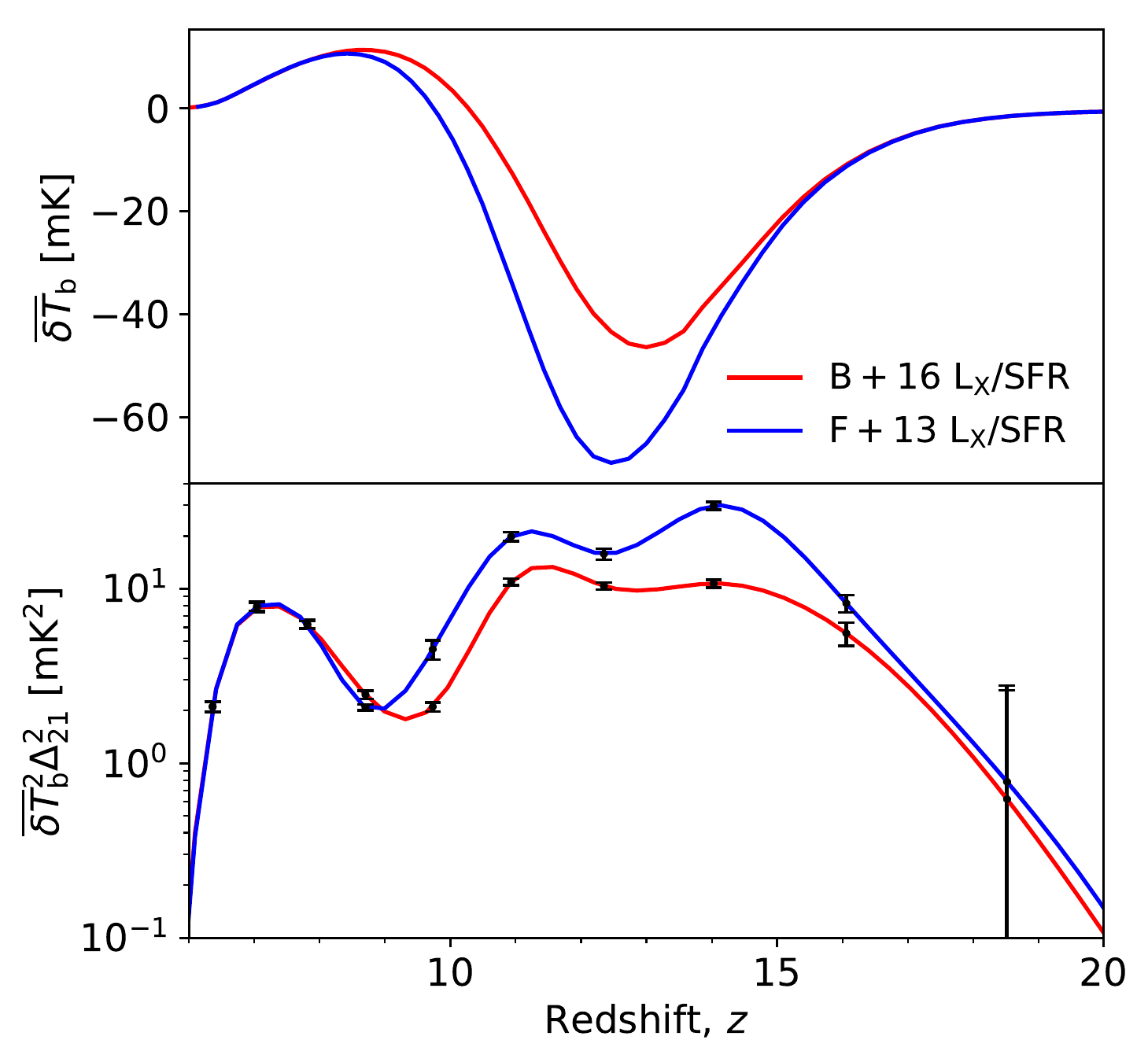}
     \caption{{\it Top panel}: redshift evolution of the global signal, $\overline{\delta T}_{\rm b}$ for \citetalias{Brorby_2016} and \citetalias{Fragos_2013a} $L_{\rm X}/\rm SFR$--$Z$ relations shown as red and blue curves, respectively. 
     \it Bottom panel: \rm corresponding redshift evolution of the 21-cm PS evaluated at $k=0.1\rm \ Mpc^{-1}$.  Error bars denote the 1$\sigma$ uncertainty expected from a 1000h observation assuming optimistic foregrounds with the Square Kilometer Array (see Sect. \ref{sec:mcmc} for details).}
    \label{fig:Tb}
\end{figure}

In the top panel of Figure \ref{fig:Tb} we plot the redshift evolution of the global 21-cm signal, $\overline{\delta T}_{\rm b}$, corresponding to the \citetalias{Brorby_2016} and \citetalias{Fragos_2013a} $L_{\rm X}/\rm SFR$--$Z$ relations with solid red and blue lines respectively.

Since the \citetalias{Brorby_2016} scaling implies stronger X-ray heating during the EoH, its $ \overline{\delta T}_{\rm b}$ has an earlier and shallower absorption trough as compared with \citetalias{Fragos_2013a}.  
The differences between the global signal evolution implied by the two $\LxSFR$--$Z$ relations peak during the epoch of heating with a 25 mK difference in the depths of the absorption troughs.

In the bottom panel of Figure  \ref{fig:Tb} we plot the redshift evolution of the 21-cm power spectrum (PS) amplitude, evaluated at $k=0.1$ Mpc$^{-1}$.  We focus on this wavemode as it corresponds to a "sweet spot"
for 21-cm interferometers: large enough to mitigate foreground contamination yet small enough to have low thermal noise (e.g. \citealt{LOFAR_2020,Trott_2020,HERA_2021a}). The redshift evolution of the large scale 21-cm power shows the characteristic three peaks, corresponding to the major astrophysical epochs: Wouthuysen-Field (\citealt{Wouthuysen_1952,Field_1958}; WF) coupling, EoH, and Epoch of Reionization (EoR).  We also show the expected thermal and cosmic variance noise for a 1000h observation, assuming optimistic foregrounds from \citet{Pober2014}, with the Square Kilometer Array (SKA; see the following section for details on how these uncertainties are calculated).

Comparing the two solid curves, we see that the \citetalias{Brorby_2016} scaling results in an earlier EoH.  This increases the overlap between the EoH and the epoch of WF coupling: IGM in regions with an under-abundance of galaxies still has different spin and kinetic temperatures at the same time that the IGM in regions with an over-abundance of galaxies becomes hot (e.g. \citealt{Mesinger_2013}).  This decreases the temperature contrast during the Cosmic Dawn, resulting in lower PS amplitudes by up to a factor of $\sim3$ for the \citetalias{Brorby_2016} scaling, compared with \citetalias{Fragos_2013a}. We note that these two scenarios can be distinguished with a 1000h observation with SKA1-low.

\section{Can constant X-ray luminosity -- SFR models reproduce the metallicity-dependent signal?}
\label{sec:mcmc}

Upcoming 21-cm observations will provide a physics-rich dataset, allowing us to {\it infer} galaxy properties from the data directly.  In particular, the signal is very sensitive to the X-ray properties of the first galaxies (e.g. \citealt{Kern_2017, Park_2019}), suggesting that the data could tell us the correct $\LxSFR$--$Z$ and MZR scalings.

However, performing inference requires parametrizing these relations and adopting physically-motivated priors on the corresponding parameters.  This is challenging, as our current understanding of stellar evolution and IMFs is insufficient to motivate such parameterizations and priors.  In principle, these difficulties can be mitigated with Bayesian model selection (e.g., \citealt{BP2019, YQ_2020b}), which we will explore in a follow-up work.  Nevertheless, having simpler models of $\LxSFR$ would make inference easier and quicker.

Indeed, most current models of the X-ray background during the Cosmic Dawn assume a constant  $\LxSFR$ (e.g. \citealt{Ghara_2020, Mondal_2020, GreigMWA_2021, GreigLOFAR_2021,HERA_2021b}; though see \citealt{Madau_2017, Eide2018}), motivated by the theoretical argument that the impact of metallicity-driven winds becomes negligible below $Z\lesssim 0.05 Z_\odot$ (e.g. \citealt{Fragos_2013a}).    
In this section, we quantify whether this simplification can recover the evolution of IGM properties of our metallicity-dependent models presented in the previous section.    We make mock 21-cm PS observations using both \citetalias{Brorby_2016} and \citetalias{Fragos_2013a} $\LxSFR$--$Z$ relations, and then perform inference assuming a constant $\LxSFR$.  As the figure of merit, we compare the recovered 21-cm power spectra and the X-ray heating rates to the ``truth'' from the mock observations.

Our two mock observations correspond to the fiducial \citetalias{Brorby_2016} and \citetalias{Fragos_2013a} models discussed in the previous section.  To compute the thermal noise, we use the python module
{\tt 21cmSENSE}\footnote{\url{https://github.com/jpober/21cmSense}} \citep{Pober2013b, Pober2014} and assume a 1000h integration (6h per night) with the SKA1-low\footnote{\url{https://astronomers.skatelescope.org/}}.  In order to tightly constrain X-ray parameters and {\it maximize the importance of the metallicity dependence}, we use the "Optimistic foreground" flag in {\tt 21cmSENSE}.  This setting assumes that the foreground wedge extends only up to the full-width half-max of the primary beam; modes outside of this contaminated wedge are assumed to be free of systematics. Our mock observations span the redshift range $z=5.8-21.6$ and wavemode range $k$= 0.1 -- 1 Mpc$^{-1}$.

For performing inference using these mock observations, we use the public {\tt 21cmMC} module\footnote{\url{ https://github.com/21cmfast/21CMMC}} \citep{Greig_2015, Greig_2017, Greig_2018}. {\tt 21cmMC} is a Bayesian sampler of \cmfast, which forward-models 21-cm light-cones.  Here we use the Multinest sampler \footnote{\url{ https://github.com/rjw57/MultiNest}} \citep{Feroz_2008,Feroze_2009}, included in \cmmc\ by \citet{YQ_2020}.

In addition to the mock 21-cm signal, we also use the following current observations in the likelihood: (i) $z\geq 6$ UV luminosity functions (\citealt{Finkelstein_2016,Ishigaki_2018,Bouwens_2021}, etc.); (ii) the electron scattering optical depth to CMB; \citealt{Planck_2016b}); and (iii) the dark fraction in the spectra of high-z quasars (\citealt{McGeer_2015}). We sample the following astrophysical parameters, adopting flat priors over the quoted ranges: 
 $\log_{10}{f_{*,10}} \in [-3,0]$, $\log_{10}f_{\rm esc, 10} \in [-3,0]$, $\alpha_{*} \in [-0.5,1]$, $\alpha_{\rm esc} \in [-1,0.5]$, $\log_{10}(M_{\rm turn}/M_{\odot}) \in [8,10]$, 
 $\log_{10}(L_{\rm X}/\rm SFR /erg\ s^{-1}M_{\odot}^{-1}yr) \in [38,44]$ and $E_{0} \in [0.1-1.5]\ \rm keV$.
 
 \begin{figure}
     \centering
    \includegraphics[width=0.5\textwidth]{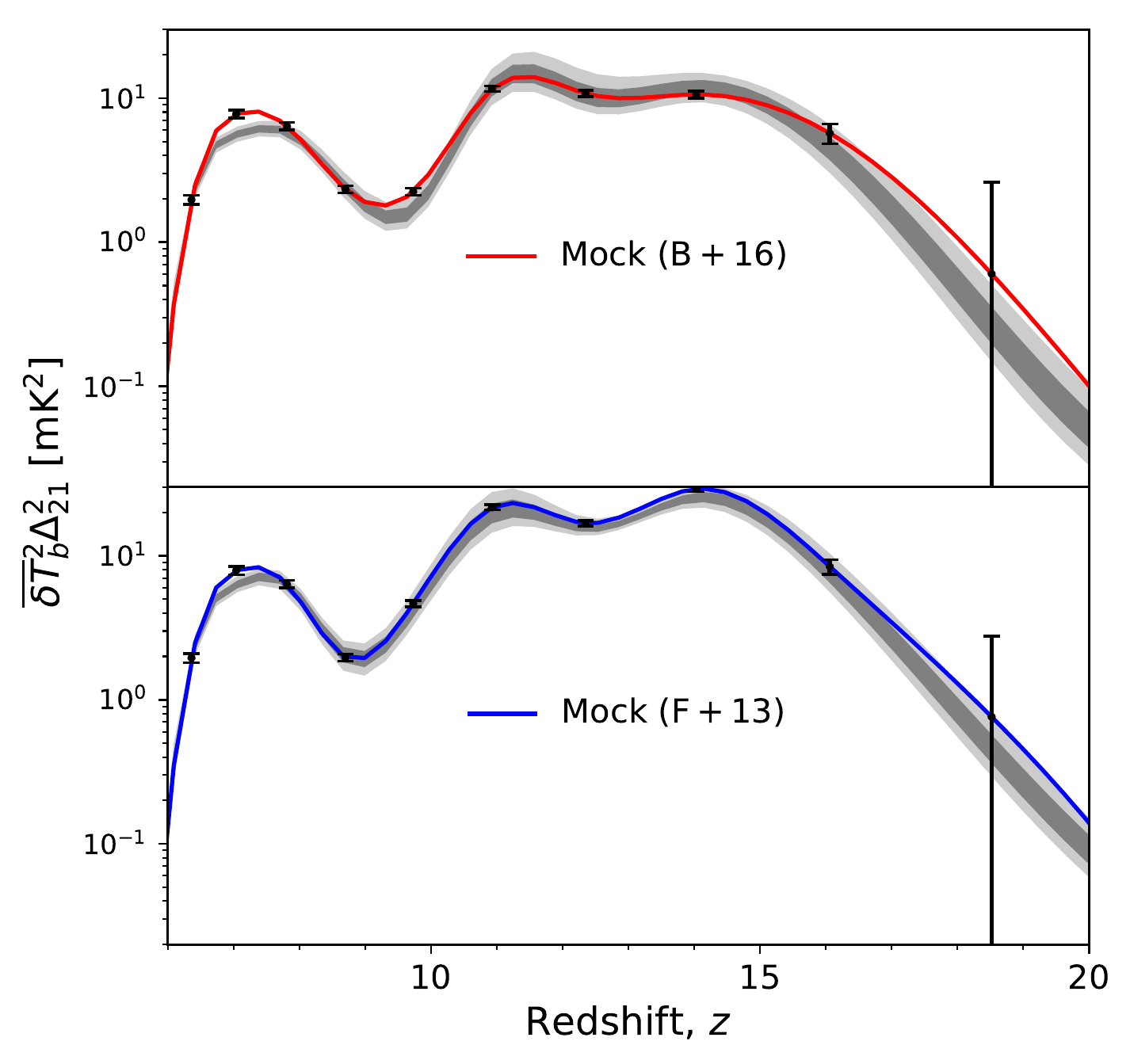}
     \caption{ Redshift evolution of 21-cm power spectra evaluated at $k=0.1 \rm Mpc^{-1}$.  The two mock observations are shown in red (top panel; \citetalias{Brorby_2016}) and blue (bottom panel; \citetalias{Fragos_2013a}), with $1\sigma$ noise (thermal + cosmic variance) denoted with error bars (see text for details.  The dark and light shaded regions correspond to  $16-84 \% $ and $2-98 \%$ credible intervals (C.I.), respectively, obtained assuming a model with a constant (i.e. metallicity independent) $\LxSFR$. 
     }
    \label{fig:PS_mcmc}
\end{figure}

In Figure \ref{fig:PS_mcmc}, we show the recovered posteriors on the evolution of large-scale ($k=0.1\ \rm Mpc^{-1}$) 21-cm PS.  Points with error bars correspond to the mock observations assuming \citetalias{Brorby_2016} (top) and \citetalias{Fragos_2013a} (bottom) scalings, while the shaded regions denote the posteriors obtained assuming a constant $\LxSFR$.  In Appendix \ref{sec:appendix}, we also show the PS posteriors as functions of wavemodes for all of the redshift samples we use in the mock observations.

The simple, constant $\LxSFR$ models are able to recover the PS evolution of the more sophisticated \citetalias{Brorby_2016} and \citetalias{Fragos_2013a} models quite well.
The posteriors are consistent with the mock data at 1$\sigma$ throughout the EoH.\footnote{We remind the reader that our choice of 'Optimistic Foregrounds' results in the small noise errors seen in this figure and in Appendix \ref{sec:appendix}.  As a result the recovered posteriors are very narrow.  We used the 'Optimistic Foregrounds' setting of {\tt 21cmSense} precisely to show the maximum expected bias from ignoring the metallicity dependence of $\LxSFR$.}

Interestingly, the largest discrepancies are found during the EoR, where the constant $\LxSFR$ models underestimate the \citetalias{Brorby_2016} 21-cm PS by up to tens of percent.  This is because most of the $\LxSFR$ constraining power comes from the EoH.  Since the  $\LxSFR$--$Z$ scaling is very steep, the EoH galaxies at $10\lesssim z \lesssim 15$ are considerably more efficient at emitting X-rays compared to EoR galaxies at $5 \lesssim z \lesssim 10$.  By fitting to the EoH, the constant $\LxSFR$ models thus end up overpredicting the X-ray background during the EoR. In the case of \citetalias{Brorby_2016}, this results in the neutral IGM patches being partially ionized by X-rays with long mean free paths.  This decreases the 21-cm contrast between ionized and neutral regions during the patchy EoR, resulting in a smaller PS amplitude. 

\begin{figure*}
    \label{fig:xHI_mcmc}
    \centering
    \includegraphics[width=0.45\textwidth]{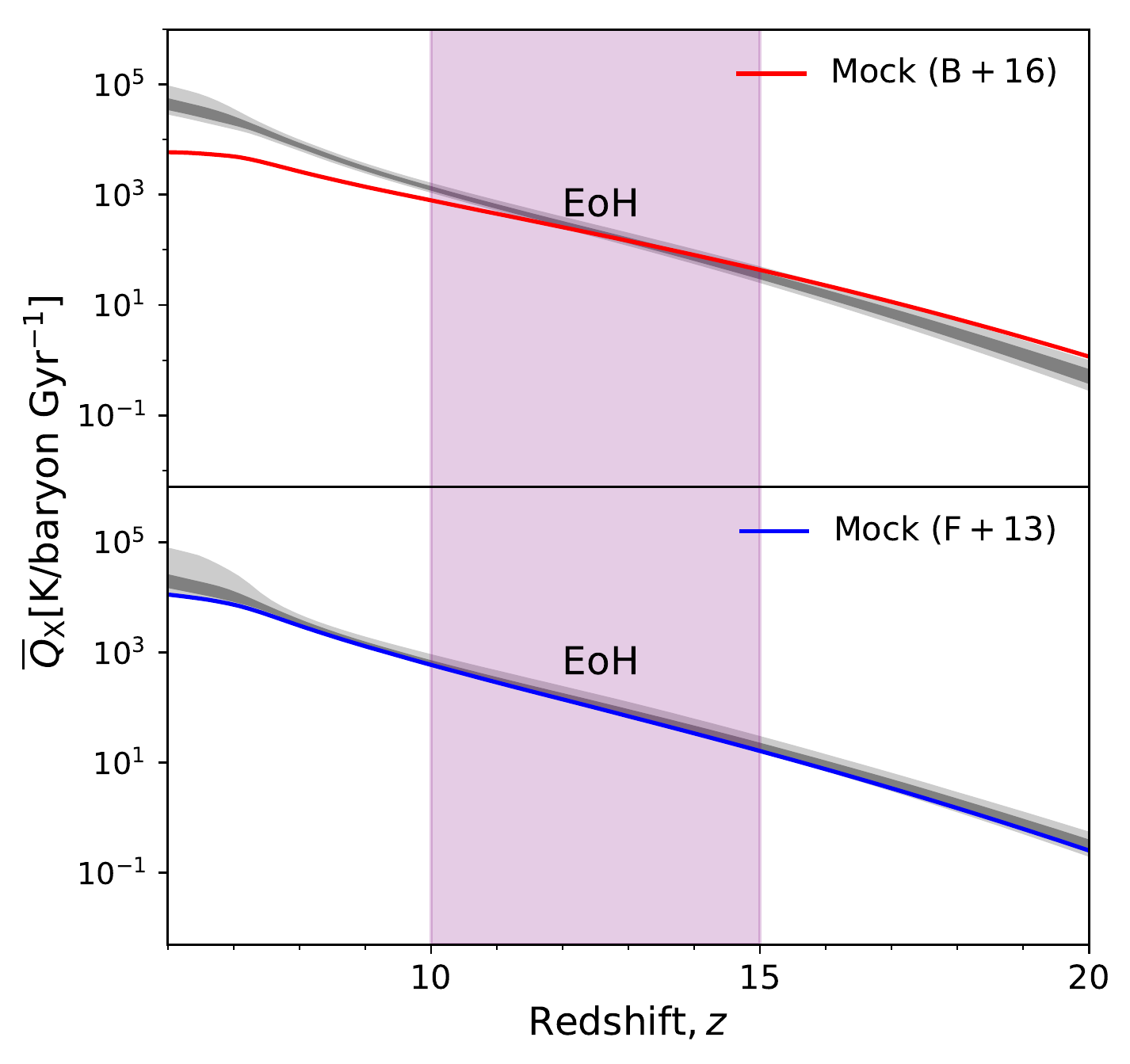}
    \includegraphics[width=0.45\textwidth]{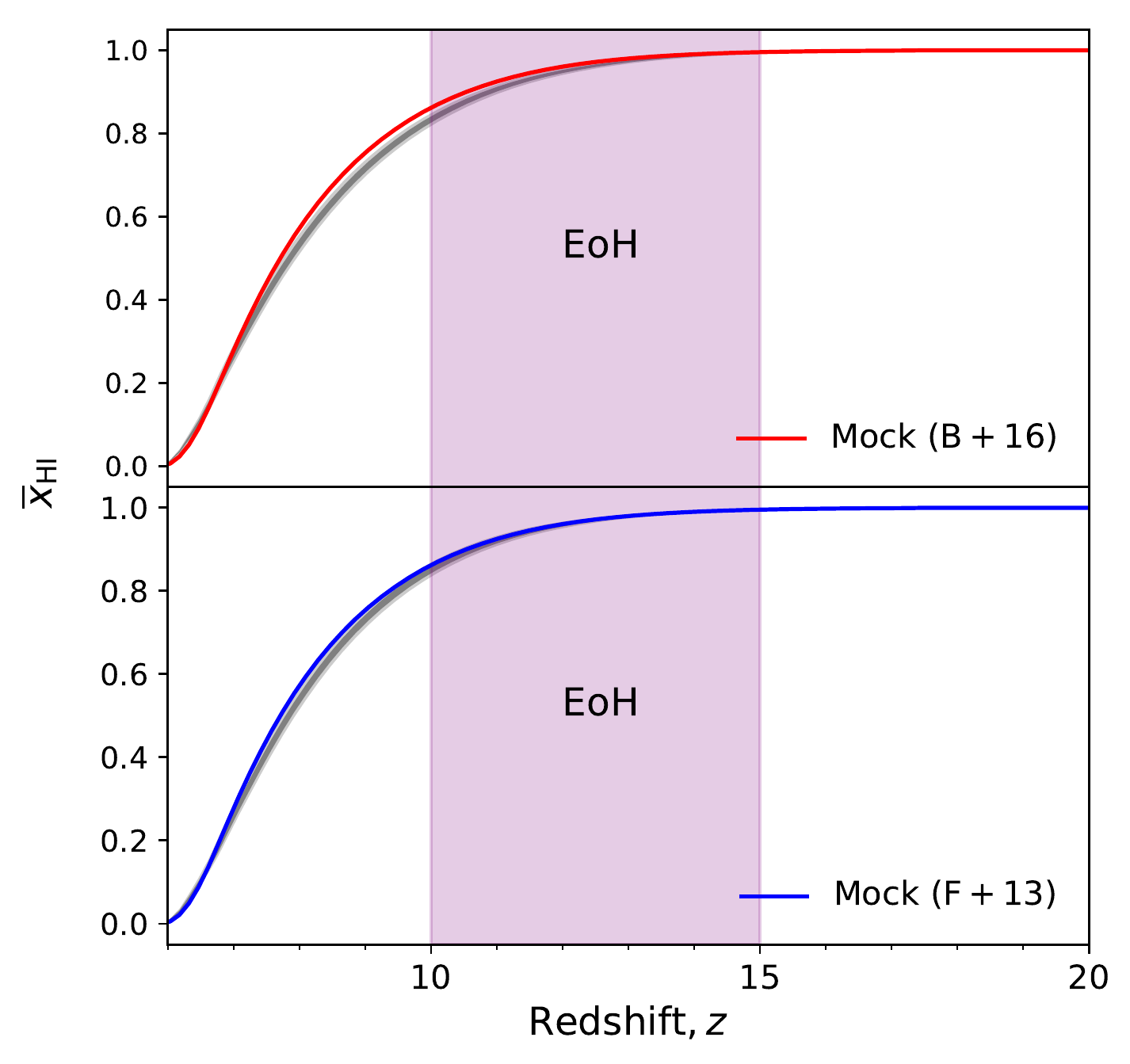}
    \caption{
Same as Figure \ref{fig:PS_mcmc}, but here showing the volume-weighted X-ray heating rate per baryon ({\it left panels}) and neutral hydrogen fraction ({\it right panels}).   
    \label{fig:Ex_mcmc}}
\end{figure*}
This is further illustrated in Figure \ref{fig:Ex_mcmc}, where we plot the analogous recovery of the X-ray heating rate ({\it left panels}) and the EoR history ({\it right panels}).  Constant $\LxSFR$ models have a steeper redshift evolution of the heating rate.  The inferred value of $\LxSFR$ corresponds to that of galaxies during the EoH at $10\lesssim z \lesssim 15$, when the 21-cm signal is most sensitive to the X-ray background.  However, towards the end of the EoR at $z\sim 6$, the inferred heating rate from constant $\LxSFR$ models can overestimate that from the mock observations by a factors of $\sim$ few -- 10.

This overestimate of the X-ray background also means that the EoR begins earlier in constant $\LxSFR$ models (see right panels of Fig. \ref{fig:Ex_mcmc}), driven by a larger contribution of X-rays to reionization.  The earlier onset of the EoR in constant $\LxSFR$ models can be partially compensated by increasing the UV ionizing emission of smaller halos.  In our model this is mostly achieved through the $\alpha_{\rm esc}$ parameter, which we find is indeed the only UV parameter whose marginal PDFs are not consistent with the value of the mock observation at more than 2$\sigma$ (see Appendix A).   We will further quantify the recovery of galaxy properties using different parametric relations for both $\LxSFR$ and MZR in future work.

Our results imply that constant $\LxSFR$ models provide a reasonable simplification for modeling the 21-cm signal during the EoH.  However, they overpredict the X-ray background at lower redshifts ($z<10$).  Tying observations over large redshift intervals without properly accounting for population-evolution (e.g. metallicity) could thus introduce significant errors.

\section{Conclusions}
\label{sec:Conclusions}

The redshifted 21-cm signal is a promising probe of heating and ionization processes in the early Universe. In particular, we expect the heating of the IGM prior to reionization to be dominated by HMXBs, hosted by galaxies too faint to be observed directly. We expect these early galaxies to be metal poor, likely implying a more X-ray luminous HMXB population than observed in local galaxies (e.g. \citealt{Linden_2010,Fragos_2013a, Basu-Zych_2013,Brorby_2016, Lehmer_2021}). 

We adopt two very different $L_{\rm X}/\rm SFR$--$Z$ relations: (i) an empirical power law from \citet{Brorby_2016}; and (ii) a theoretical stellar evolution model from \citet{Fragos_2013a}.  For the same galaxy evolution model, (i) and (ii) result in differences of up to a factor of $\sim$2--3 in the IGM temperature and 21-cm signal during the Cosmic Dawn.

HMXBs hosted by galaxies with SFRs of order $10^{-3}$ -- $10^{-1}$ $M_\odot$ yr$^{-1}$ and metallicities of order $0.001$ -- $0.01$ $Z_\odot$ dominate the IGM heating.  The assumed $L_{\rm X}/\rm SFR$--$Z$ relation can shift these ranges by factors of $\sim$ few.

We also use the two $\LxSFR$--$Z$ relations to compute mock 21-cm PS observations, assuming optimistic foreground removal and 1000h integration with SKA.  We performed inference using the common simplification of a constant $\LxSFR$.  The constant $\LxSFR$ models reproduce the IGM properties from the metallicity-dependent $\LxSFR$ simulations quite well.  However, since the inferred value of $\LxSFR$ corresponds to the dominant population during the EoH ($z\sim10$--15; when the 21-cm signal is most sensitive to the IGM temperature), they overpredict the XRB at lower redshifts ($z\lesssim10$).  Thus, accurate inference over a broad range of redshifts should account for metallicity evolution and the $\LxSFR$--$Z$ relation.

\section*{Acknowledgements}
We gratefully acknowledge computational resources of the Center for High Performance Computing (CHPC) at Scuola Normale Superiore (SNS). YQ acknowledges that part of this work was supported by the Australian Research Council Centre of Excellence for All Sky Astrophysics in 3 Dimensions (ASTRO 3D), through project \#CE170100013; and some of the simulations presented in this work were run on the OzSTAR national facility at Swinburne University of Technology. AP acknowledges support from the ERC Advanced Grant INTERSTELLAR H2020/740120. TF acknowledges  support from the Swiss National Science Foundation Professorship grant (project number PP00P2\_176868) AB acknowledges support by NASA under award number 80GSFC21M0002. 

\section{Data availability} 
The data generated during this work is available from the corresponding author on request.

\bibliography{Xrays}

\appendix
\label{sec:appendix}
\section{21-cm power spectra}
Here we show the 21-cm power spectra used in the two inferences from this paper.  In Figure \ref{fig:Brorby_posterior}, the black curves correspond to the \citetalias{Brorby_2016} $\LxSFR$--$Z$ relation, with error bars marking 1$\sigma$ noise.  The recovered posterior assuming a metallicity-independent $\LxSFR$ is shown in red.  Figure Fig. \ref{fig:Fragos_posterior} is analogous to Fig. \ref{fig:Brorby_posterior}, but using the \citetalias{Fragos_2013a} relation for the mock observation.  We see that the constant $\LxSFR$ models reproduce the 21-cm PS quite well over the EoH.  However, the steeper redshift evolution implied by the constant $\LxSFR$ models over-estimates the contribution of X-rays to the early stages of the EoR (see Fig. \ref{fig:xHI_mcmc}), which translates to a $\sim$ tens of percent underprediction of large scale 21-cm power (c.f. Fig. \ref{fig:PS_mcmc}) and a biased recovery of the ionizing escape fraction scaling with halo mass, $\alpha_{\rm esc}$, shown in Fig. \ref{fig:alphax}.  We explore the recovery of galaxy parameters using different parametric relations for $\LxSFR$ and MZR in a follow-up work. 

\renewcommand{\thefigure}{A\arabic{figure}}
\setcounter{figure}{1}
\begin{figure*}
    \centering
    \includegraphics[scale=0.7]{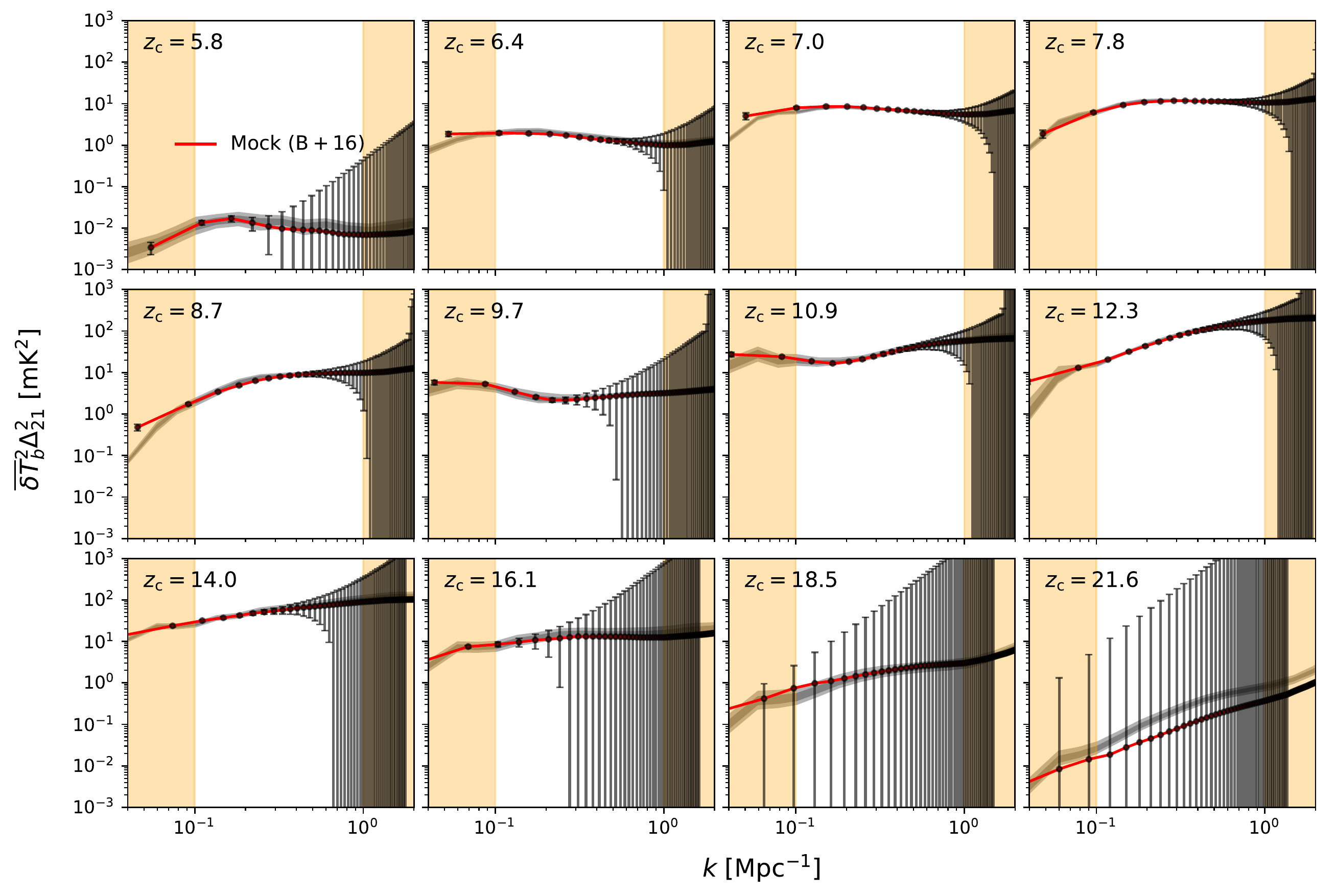}
    \caption{21-cm power spectra. The mock observation corresponding to the \citetalias{Brorby_2016} $\LxSFR$--$Z$ scaling is shown in red with error bars representing  $1\sigma$ noise assuming a 1000 hr observation with SKA1-low. The dark and light shaded regions denote 16-84\% and 2-98\% confidence intervals, respectively, obtained assuming a constant $\LxSFR$. $z_{\rm c}$ denotes the central redshift of the light-cone chunk used to compute the power spectra.  We demarcate with yellow stripes the $k$ modes outside of the $k$ = 0.1 -- 1 Mpc$^{-1}$ range used to compute the likelihood.
    \label{fig:Brorby_posterior}}
\end{figure*}
\begin{figure*}
    \centering
    \includegraphics[scale=0.7]{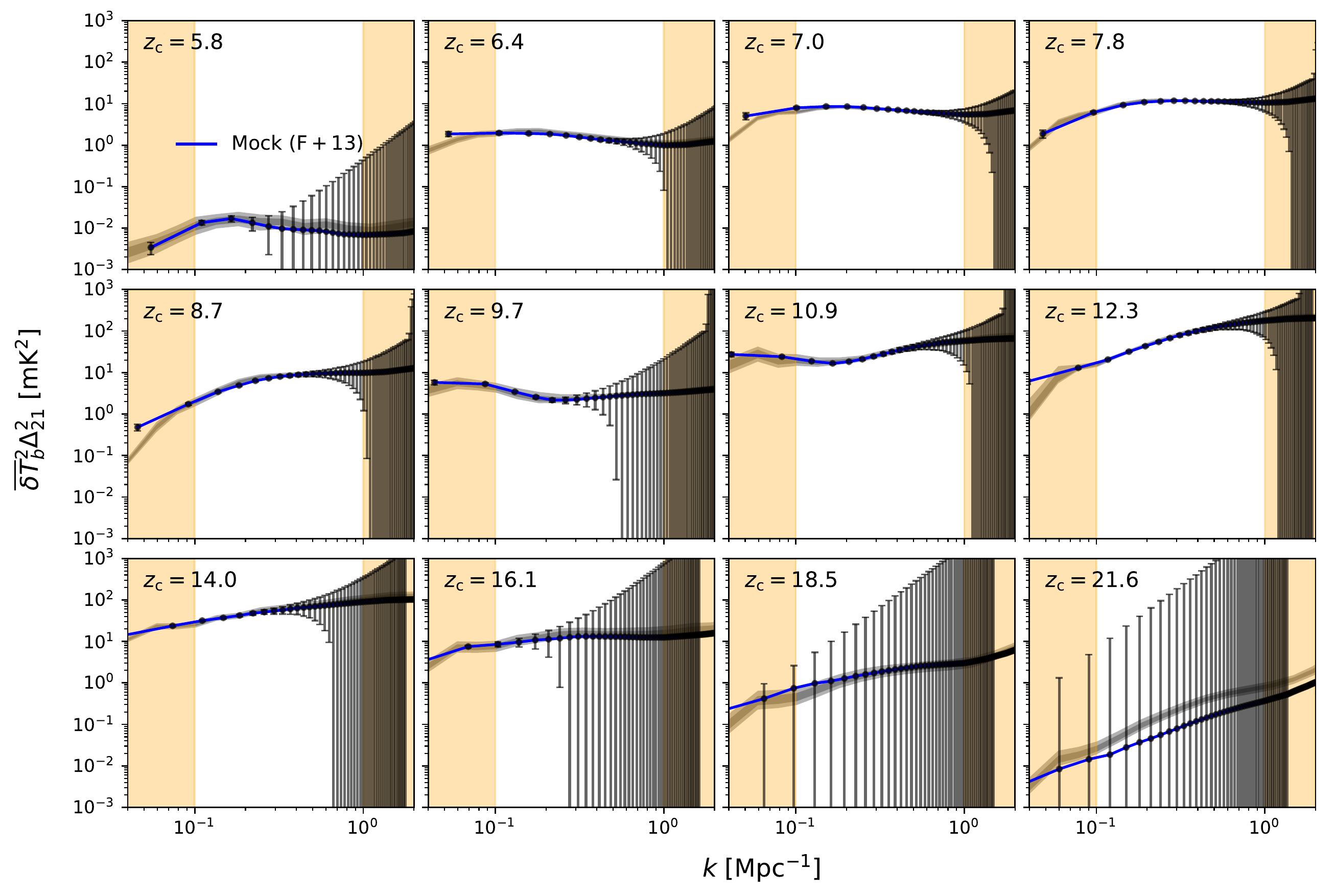}
    \caption{Same as Figure \ref{fig:Brorby_posterior} but using \citetalias{Fragos_2013a} $\LxSFR$--$Z$ scaling as our mock observation (blue curve).
    \label{fig:Fragos_posterior}}
\end{figure*}

\renewcommand{\thefigure}{A\arabic{figure}}
\begin{figure}
    \centering
    \includegraphics[width=0.5\textwidth]{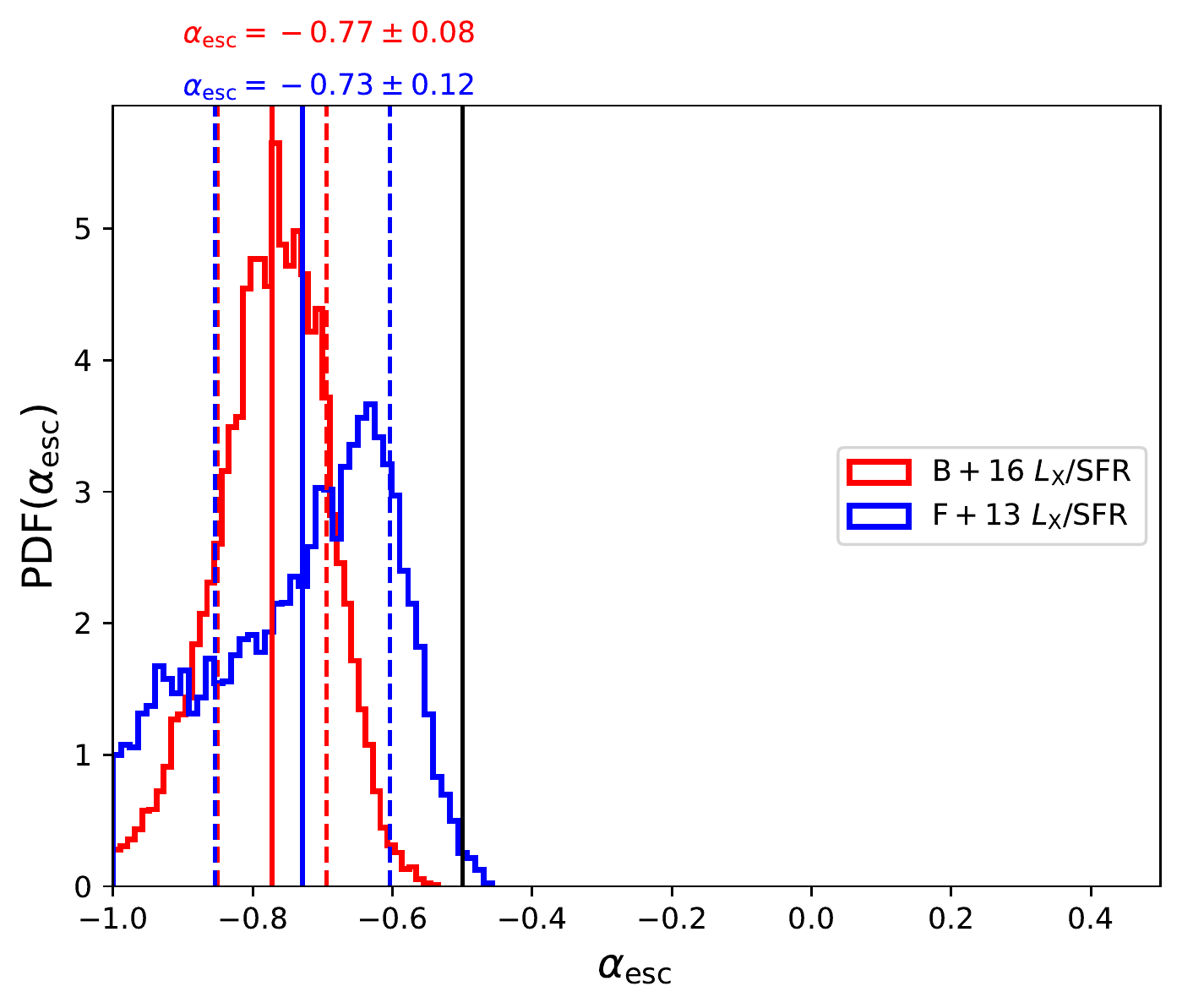}
    \caption{Marginalized 1D PDFs of  the power law scaling index of the ionizing escape fraction with halo mass, $\alpha_{\rm esc}$, for constant $\LxSFR$ models.  The PDFs inferred from mock data created using \citetalias{Brorby_2016} / \citetalias{Fragos_2013a} relations are denoted in red / blue.  The mean and 1 $\sigma$ r.m.s. width are denoted with solid and dashed vertical lines, respectively.  The vertical black denotes the "true" value used in making the mock simulations. The horizontal range shows the extent of our flat prior on $\alpha_{\rm esc}$.  Due to our choice of "Optimistic foregrounds", $\alpha_{\rm esc}$ is very tightly constrained from the mock data.  However, the recovered values for the constant $\LxSFR$ models are biased, in order to compensate for the their implied additional contribution of X-rays to the very early stages of reionization, as discussed in the text.  We will return to the recovery of galaxy properties in a follow-up work, including parametric models for the X-ray properties of the first galaxies.
    }
    \label{fig:alphax}
\end{figure}

\end{document}